\documentclass[]{aa}

\usepackage{graphicx}
\usepackage{txfonts}
\usepackage{natbib}

\newcommand{\un}[1]{{\,{\rm #1}}}
\newcommand{\dw}[1]{_{\scriptstyle \mathrm{#1}}}

\begin{document}

\title{Impact of non-uniform surface magnetic fields on stellar winds}
\titlerunning{Latitude-dependent magnetised winds}

\author{V Holzwarth}
\offprints{V Holzwarth, \email{vrh1@st-andrews.ac.uk}}

\institute{School of Physics and Astronomy, University of St Andrews, 
North Haugh, St Andrews, Fife KY16 9SS, Scotland}

\date{Received ; accepted}

\abstract{
Observations of active stars reveal highly non-uniform surface 
distributions of magnetic flux.
Theoretical models considering magnetised stellar winds however often 
presume uniform surface magnetic fields, characterised by a single
magnetic field strength.
The present work investigates the impact of non-uniform
surface magnetic field distributions on the stellar mass and
angular momentum loss rates as well as on the effective Alfv\'enic
radius of the wind.
Assuming an axial symmetric and polytropic magnetised wind, the 
approach of \citet{1967ApJ...148..217W} is extended to non-equatorial 
latitudes to quantify the impact of latitude-dependent magnetic field
distributions over a large range of stellar rotation rates and thermal
wind properties.
Motivated by recent observational results, the analytically prescribed 
field patterns are dominated by magnetic flux concentrations at 
intermediate and high latitudes.
The global stellar mass loss rates are found to be rather insensitive
to non-uniformities of the surface magnetic field.
Depending on the non-uniformity of the field distribution, the angular 
momentum loss rates deviate in contrast at all rotation rates between 
$-60\%$ and $10\%$ from the \citeauthor{1967ApJ...148..217W}-values, 
and the effective Alfv\'enic radii up to about $\pm 25\%$.
These large variations albeit equal amounts of total magnetic flux
indicate that a classification of stellar surface magnetic
fields through a single field strength is insufficient, and that their
non-uniformity has to be taken into account.
The consequences for applications involving magnetised stellar winds 
are discussed in view of the rotational evolution of solar-like stars 
and of the observational determination of their mass loss rates using
the terminal velocity and ram pressure of the wind.
For rapidly rotating stars the latitudinal variation of the wind ram
pressure is found to exceed, depending on the actual field distribution
on the stellar surface, over two orders of magnitude.
The assumption of a spherical symmetric wind geometry may therefore
lead to a significant over- or underestimation of the stellar mass
loss rate.
}

\maketitle

\section{Introduction}
\label{intro}
Stars with hot coronae are subject to mass and angular momentum losses 
due to stellar winds \citep{1958ApJ...128..664P}.
In the presence of magnetic fields the angular momentum (AM) loss is
significantly enhanced, because the escaping ionised plasma is forced
into co-rotation with the star out to distances much larger than the
stellar radius \citep{1962AnAp...25...18S}.
Further to its contribution to the specific AM of the plasma, the 
stress of bent magnetic field lines adds to the acceleration of the
wind by magneto-centrifugal driving.
In rapidly rotating stars this inherently latitude-dependent mechanism
is expected to dominate over the isotropic thermal driving
\citep{1969ApJ...158..727M}.

Whereas early studies of stellar winds assumed uniform or dipolar
surface magnetic fields, high-resolution observations of the solar
atmosphere now indicate much more complex boundary conditions at the
base of the wind.
Doppler Imaging (DI) observations of active stars show non-uniform 
surface brightness distributions, which are ascribed to the presence of
magnetic flux in the form of dark spots \citep{2001astr.conf..183C}.
They show that on stars rotating more rapidly than the Sun magnetic 
flux is not only located in equatorial regions, but also at 
intermediate and high latitudes \citep[][ and references
therein]{2002AN....323..309S}.
Although starspots can a priori not uniquely be associated with wind
loss regions, they nevertheless identify locations of possibly open 
magnetic field structures.
Whereas spectroscopic DI enables the reconstruction of starspot maps, 
spectro-polarimetric Zeeman-DI directly confirms the magnetic origin of
the dark surface features \citep{1997MNRAS.291..658D}.
In combination with field extrapolation techniques, the observed 
surface magnetic field distributions serve as boundary conditions for
the reconstruction of coronal magnetic field topologies and thus for
the determination of both closed and open field regions
\citep[e.g.][]{2002MNRAS.333..339J}.
Recent results show that on rapidly rotating stars open magnetic flux
is organised in belt-like structures mainly between intermediate and
high latitudes \citep{2004MNRAS.355.1066M}.

\citet[ hereafter WD]{1967ApJ...148..217W} theoretically investigated 
the radial structure and properties of the magnetised solar wind
assuming that the wind structure determined in the equatorial plane is 
representative for the entire sphere, implying a monopolar, spherical
symmetric field geometry.
Parameter studies have been carried out in the framework of this model 
to investigate the dependence of the AM loss rate on the characteristic
magnetic field strength, the stellar rotation rate, and thermal wind 
properties \citep{1976ApJ...210..498B, 1976ApJ...206..768Y}.
\citet{1968MNRAS.138..359M} and \citet{1987MNRAS.226...57M} considered 
dipolar surface fields with closed magnetic structures forming `dead 
zones' at the equator and radial open fields lines further away from 
the star; \citet{1974MNRAS.166..683O} extended this approach to 
different analytical poloidal field structures.
A self-consistent determination of the poloidal field structure
requires the solution of non-linear partial differential equations,
which represents a formidable aspect of the overall wind problem.
Theoretical and numerical investigations including this aspect use
semi-analytical techniques based on the separation of variables
\citep{1998MNRAS.298..777V, 2001A&A...371..240L, 2002A&A...389.1068S}
or multi-dimensional MHD simulations \citep{1985A&A...152..121S,
1999A&A...343..251K, 2000ApJ...530.1036K}, respectively.
They reveal at large distances from the star a rotation-dependent
latitudinal deflection of the magnetic field lines toward the symmetry
axis as a result of the collimating effect of the azimuthal field
component \citep[e.g.][]{1989ApJ...347.1055H, 2000A&A...356..989T,
2000MNRAS.318..250O}.
The self-consistent treatments are mainly focused on the (asymptotic)
wind structure without going into details about mass and AM losses for
different stellar parameters.
Owing to the non-linearity of the influence of the poloidal 
field component on the overall wind structure, the differences between 
self-consistent methods and simplified approximations are difficult to 
predict and have to be verified through direct comparisons. 
Therefore, the results of WD-like approaches should a priori be
regarded as estimates within the framework of the applied
simplifications.

The approaches of \citet{1967ApJ...148..217W} and 
\citet{1987MNRAS.226...57M} are widely used for the determination of
AM loss rates in the course of the rotational evolution of stars
\citep{1981ApJ...243..625E, 1994MNRAS.269.1099C, 1999MNRAS.302..203L}.
In contrast, \citet{1997A&A...325.1039S} and \citet{2005srestnsdoomf} 
used specific modifications of the WD model to investigate the impact
of non-uniform surface fields.
To determine mass loss rates of solar-like stars,
\citet{1998ApJ...492..788W} and \citet{2002ApJ...574..412W} devised a
method using the line-of-sight absorption of the Ly$\alpha$ emission 
by surrounding astrospheres, which are formed by the collision between
the (magnetised) wind from the star and the local interstellar medium.
The size of the astrosphere as well as the amount of absorption depend 
on the ram pressure of the wind, which is a function of the stellar
mass loss and wind velocity. 
For the determination of the total stellar mass loss rate they assume a
constant (i.e. spherical symmetric and independent of the rotation rate
and latitude) wind velocity.
Based on their results, they formulate an empirical relationship 
between mass loss rates and coronal X-ray fluxes of solar-like stars.

Motivated by the progress in the determination of stellar surface 
features and mass loss rates, the present work applies the WD 
wind model to non-equatorial latitudes to investigate the influence of
latitude-dependent surface magnetic fields on the total mass and AM
loss rates of cool stars.
In contrast to self-consistent treatments of the magnetised wind 
problem, which are focused on detailed properties of the wind 
structure, the main objective here is the global qualification and 
quantification of the influence of observed surface magnetic field 
distributions in relation to the originally spherical symmetric WD 
method.
The applied wind model represents an efficient way to include the 
aspect of non-uniform surface magnetic fields in the rotational 
evolution of solar-like stars and in the observational determination of
their mass loss rates.

\section{Model assumptions}
\label{moas}
The stellar mass and AM loss rates through latitude-dependent 
magnetised winds are determined for a solar-like star by applying the
WD approach to open magnetic field lines at all latitudes separately
(cf.~Appendix \ref{powi}).
The poloidal components of the magnetic field and the flow velocity are
assumed to be radial, neglecting the trans-field component of the
(stationary) equation of motion perpendicular to the magnetic field.
Retaining the full azimuthal wind dependency, the open field lines form
spirals on coni with constant opening angles, whose tips are located
in the stellar centre.

The polytropic wind is assumed to be symmetric with respect to both the 
rotation axis and the equator, enabling a limitation of the analyses to
the range of co-latitudes $0< \theta \le \pi/2$.
The wind solutions are determined through the stellar rotation rate and
the magnetic and thermal wind conditions prescribed at a reference
level, $r_0= 1.1\un{\mathrm{R_\odot}}$, close to the stellar surface.
The uniform rotation rates cover the range $\Omega= 10^{-6} - 
5.4\cdot10^{-4}\un{s^{-1}}\approx 0.3 - 180\un{\Omega_\odot}$, that is 
rotation periods $73\un{d} - 3\un{h}$; the upper limit of the rotation
rates corresponds to the case when the co-rotation radius,
Eq.~(\ref{rdeldef}), in the equatorial plane is equal to the reference
radius.

\subsection{Thermal and magnetic wind properties}
The wind temperature and (particle) density at the reference level are
approximated through solar-like values, that is $T_0= 2\cdot10^6\un{K}$
and $n_0= 10^8\un{cm^{-3}}$, respectively.
With a mean atomic weight $\mu= \mu_{\odot}\simeq 0.6$, these values 
imply a gas density of $\rho_0\simeq 10^{-16}\un{g/cm^{3}}$ and a gas 
pressure of $p_0\simeq 3\cdot10^{-2}\un{dyn/cm^2}$.
The entropy change of the wind with increasing distance from the star
is quantified through the polytropic index, $\Gamma= 1.15$.
These values fall in the range of values of similar studies (e.g.,
$\Gamma\simeq 1.2$ and $T_0= 2.7\cdot10^6\un{K}$,
\citealt{1967ApJ...148..217W, 1985A&A...152..121S}; $\Gamma= 1.13$ and
$T_0= 1.5\cdot10^6\un{K}$, \citealt{1995A&A...294..469K}), and have
been chosen to ensure that even plasma emanating at high latitudes,
where the magneto-centrifugal driving is inherently less efficient,
escapes from the star owing to a sufficiently strong thermal driving.
The thermal boundary conditions are taken to be independent of both
latitude and rotation rate.

The magnetic wind properties are determined through the radial magnetic
field component at the reference level.
For the uniform Constant Field (\emph{CO}) distributions the field 
strengths are taken to be $B_{r,0}= 3, 30$, and $300\un{G}$,
corresponding to an (unsigned) open magnetic flux between $\Phi\simeq
2\cdot10^{23-25}\un{Mx}$.
The lower limit is of the order of the average solar magnetic field 
strength, whereas the upper limit is in accordance with
spectro-polarimetric observations of rapidly rotating stars, which
indicate the existence of field strengths up to two orders of magnitude
larger than in the case of the Sun \citep{1997MNRAS.291....1D}.

\begin{figure}
\includegraphics[width=\hsize]{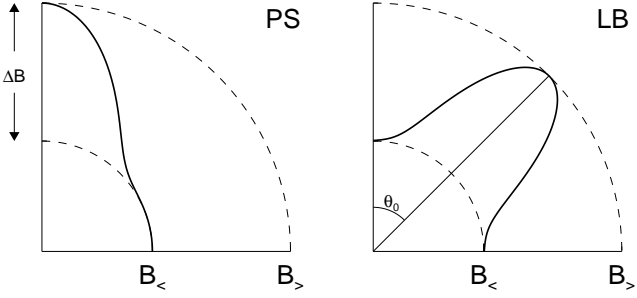}
\caption{ Polar Spot (\emph{PS}) and Latitudinal Belt (\emph{LB}) 
distributions of the radial magnetic field component, $B_{r,0} 
(\theta)$, at the reference level. }
\label{bmodels.fig}
\end{figure}

In the case of non-uniform magnetic fields, latitude-dependent 
distributions are superposed on a constant background field of $B_<=
3\un{G}$ (Fig.~\ref{bmodels.fig}).
The peak field strength, $B_> = B_< + \Delta B$, is chosen so that the 
\emph{surface averaged} magnetic field strength,
\begin{equation}
\bar{B}_{r,0} 
= 
\int \limits_0^{\pi/2} B_{r,0} \left( \theta \right) \sin \theta\, 
d\theta
\ ,
\label{defbr0}
\end{equation}
of each distribution is $30\un{G}$, to allow for a comparison with the
intermediate case of the Constant Field distributions above.
Using the analytical form
\begin{equation}
B_{r,0} \left( \theta \right) 
= 
B_< + \Delta B \cdot \cos^n \left( \theta - \theta_0 \right)
\ ,
\label{defbrtheta}
\end{equation}
the latitude-dependent magnetic field distributions considered here are
the
\begin{enumerate}
\item 
{\sf Polar Spot} model (PS), with $n= 8, \theta_0= 0$, and $\Delta B= 
243\un{G}$, and the
\item 
{\sf Latitudinal Belt} model (LB), with $n= 16, \theta_0= 45\degr$, and
$\Delta B= 63.8\un{G}$.
\end{enumerate}
The first distribution is motivated by observationally determined 
surface brightness maps of active stars \citep{2002AN....323..309S}, 
which frequently show spot concentrations at high latitudes.
The second one is initiated through recent results of coronal magnetic
field extrapolations based on Zeeman-Doppler Imaging observations of
rapid rotators, which reveal a gathering of open magnetic flux also at
intermediate latitudes \citep{2004MNRAS.355.1066M}.

\subsection{Stellar mass and angular momentum loss rates}
The contribution of a plasma stream to the stellar mass loss rate is
given by
\begin{equation}
d\dot{M}
=
F\dw{M} d\sigma
=
2 \pi r\dw{A}^2 \rho\dw{A} v_{r,{\rm A}} \sin \theta\, d\theta
\ ,
\label{tmalpl}
\end{equation}
with $d\sigma= \sin\theta\, d\theta\,d\phi$ being the solid angle 
occupied by the escaping plasma flow. 
The mass flux per solid angle, $F\dw{M}= \rho v_r r^2= \rho\dw{A}
v_{r,{\rm A}} r\dw{A}^2$, is constant along individual field lines but 
latitude-dependent, since the plasma density, $\rho\dw{A}$, and radial 
flow velocity, $v_{r,{\rm A}}$, at the Alfv\'enic point, $r\dw{A}$, are
functions of the co-latitude.
The total stellar mass loss rate is given by
\begin{equation}
\dot{M}
=
\dot{M}\dw{WD} 
\int \limits_0^{\pi/2} \left( \frac{r\dw{A}}{\bar{r}\dw{A}} \right)^2 
\left( \frac{\rho\dw{A}}{\bar{\rho}\dw{A}} \right)
\left( \frac{v_{r,{\rm A}}}{\bar{v}_{r,{\rm A}}} \right)
\sin \theta\, d \theta
\ .
\label{tmal}
\end{equation}
The WD mass loss rate,
\begin{equation}
\dot{M}\dw{WD}
=
4 \pi \bar{r}\dw{A}^2 \bar{\rho}\dw{A} \bar{v}_{r,{\rm A}}
\ ,
\label{tmalwd}
\end{equation}
is derived from the assumption that the density, $\bar{\rho}\dw{A}$,
and radial flow velocity, $\bar{v}_{r,{\rm A}}$, at the Alfv\'enic
point, $\bar{r}\dw{A}$, \emph{in the equatorial plane} are
representative for the wind structure at all latitudes, implying
$r\dw{A} (\theta)= \bar{r}\dw{A}, \rho\dw{A} (\theta)=
\bar{\rho}\dw{A}$, and $v_{r,{\rm A}} (\theta)= \bar{v}_{r,{\rm A}}$.
All WD values are calculated using the surface averaged field strength,
Eq.~(\ref{defbr0}); the determination of wind solutions is outlined in
Appendix \ref{powi}.
  
The AM momentum loss per latitude is
\begin{equation}
d\dot{J}
=
L\, F\dw{M} d\sigma
=
2 \pi \Omega r\dw{A}^4 \rho\dw{A} v_{r,{\rm A}} \sin^3 \theta\, d\theta
\ ,
\label{tamlpl}
\end{equation}
where $L= \Omega \left( r\dw{A} \sin \theta \right)^2$ is the specific 
AM along an individual field line [cf.\ Eqs.~(\ref{deflco}) \&
(\ref{lcoconst})].
The total AM loss rate of the star is then given by
\begin{eqnarray}
\dot{J}
& = &
\dot{J}\dw{WD} \frac{3}{2}
\int \limits_0^{\pi/2} \left( \frac{r\dw{A}}{\bar{r}\dw{A}} \right)^4 
\left( \frac{\rho\dw{A}}{\bar{\rho}\dw{A}} \right)
\left( \frac{v_{r,{\rm A}}}{\bar{v}_{r,{\rm A}}} \right)
\sin^3 \theta\, d \theta
\ ,
\label{taml}
\end{eqnarray}
with the respective WD value 
\begin{equation}
\dot{J}\dw{WD}
=
\frac{8\pi}{3} \Omega \bar{r}\dw{A}^4 \bar{\rho}\dw{A} 
\bar{v}_{r,{\rm A}}
= 
\frac{2}{3} \Omega \bar{r}\dw{A}^2 \dot{M}\dw{WD}
\ .
\label{tamlwd}
\end{equation}
Latitude-dependent boundary conditions at the base of the stellar wind
result in quantities $(r\dw{A}, \rho\dw{A}, v_{r,{\rm A}})$, which are 
functions of the co-latitude and a priori different from 
$(\bar{r}\dw{A}, \bar{\rho}\dw{A}, \bar{v}_{r,{\rm A}})$ in the 
equatorial plane.
Even boundary conditions constant at all latitudes cause a deviation of
the wind structure from spherical symmetry, since the contribution of
the magneto-centrifugal driving to the overall wind acceleration along
non-equatorial field lines is smaller than in the equatorial plane.

Studies considering the rotational evolution of stars often determine 
AM loss rates not through explicit boundary conditions, but by using 
Eq.~(\ref{tamlwd}) in combination with either prescribed or empirical
mass loss rates \citep[e.g.,][]{1988ApJ...333..236K,
2002ApJ...574..412W}, deriving the required Alfv\'enic radius by more
qualitative arguments and/or relationships.
The influence of non-uniform field distributions on this approach is
quantified through the \emph{effective} Alfv\'enic radius,
\begin{equation}
r\dw{A,eff}
=
\bar{r}\dw{A} \left( 
 \frac{ \dot{J} / \dot{J}\dw{WD} }{ \dot{M} / \dot{M}\dw{WD} } 
\right)^{1/2}
\ ,
\label{defraeff}
\end{equation}
which describes the functional relation between the latitude-integrated
mass and AM loss rates in comparison with the WD results.

\section{Results}
\label{resu}

\subsection{Reference cases}
\label{refa}
\begin{figure}
\includegraphics[width=\hsize]{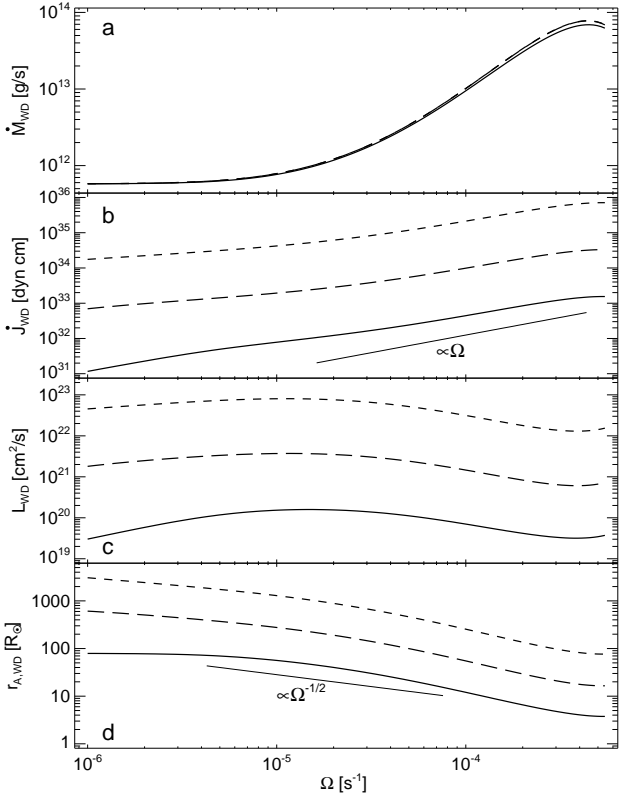}
\caption{ Reference values determined following the approach of 
\citet{1967ApJ...148..217W}, with $\bar{B}_{r,0}= 3\un{G}$
(\emph{solid}), $30\un{G}$ (\emph{long dashed}), and $300\un{G}$
(\emph{short dashed}).
Panel \textbf{a}: Mass loss rate, $\dot{M}\dw{WD}$; for the two larger
field strengths the mass loss rates are quite similar and hardly
distinguishable.
Panel \textbf{b}: Angular momentum loss rate, $\dot{J}\dw{WD}$; the
thin line indicates the slope of a function linear in the stellar
rotation rate, $\Omega$.
Panel \textbf{c}: Specific angular momentum, $L\dw{WD}$.
Panel \textbf{d}: Alfv\'enic radius, $r\dw{WD}$; the thin line
indicates the slope of a function $\propto \Omega^{-1/2}$. }
\label{wdref.fig}
\end{figure}

The mass loss rates, $\dot{M}\dw{WD}$, AM loss rates, $\dot{J}\dw{WD}$, 
and Alfv\'enic radii, $\bar{r}\dw{A}$, determined in the equatorial 
plane following the approach of \citet{1967ApJ...148..217W}, are used
as reference values.
The stellar mass loss rate is controlled by the wind structure in the 
subsonic flow regime close to the star.
Its value depends significantly on the stellar rotation rate but hardly
on the magnetic field strength (Fig.~\ref{wdref.fig}a).
The increase of the AM loss rate with the rotation rate 
(Fig.~\ref{wdref.fig}b) is due to the strong increase of the mass
loss rate, which outbalances minor variations of the specific AM 
(Fig.~\ref{wdref.fig}c).
Approximations based on rotation-independent magnetic field strengths 
imply a linear increase of the AM loss rate, that is $\dot{J}\propto 
\Omega^a$ with $a\simeq 1$, whereas the functional relations here are
found to be sub-linear.
For rotation rates $\Omega\gtrsim 2\cdot10^{-5}\un{s^{-1}}\simeq
7\un{\Omega_\odot}$ a curve fit yields $a\simeq 0.8$.
Below this rotation rate $a\simeq 0.8\ (3\un{G}), 0.45\ (30\un{G})$ and 
$0.4\ (300\un{G})$, respectively.
Due to its explicit linear dependence on the stellar rotation rate, the
specific AM increases at slow rotation rates, where the decline of
the square of the Alfv\'enic radius is sub-linear
(Fig.~\ref{wdref.fig}d).
At higher rotation rates this decline becomes super-linear and the 
specific AM consequently smaller, entailing a maximum at 
intermediate rotation rates.

To verify the impact of non-uniform field distributions with respect to
the spherical symmetric WD approach, in the following the results are 
mainly given in terms of relative deviations, $\delta F= \left( F - 
F\dw{WD} \right) / F\dw{WD}$, where $F$ is the mass loss rate, 
Eq.~(\ref{tmal}), the AM loss rate, Eq.~(\ref{taml}), or the 
(effective) Alfv\'enic radius, Eq.~(\ref{defraeff}), respectively.
Further quantities have been used to normalise the latitudinal profiles
in Figs.~\ref{mdotPL.fig}, \ref{jdotPL.fig}, and \ref{ramp.fig}; their
values are given in Table \ref{wdrefvalues}.
\begin{table}
\caption{Reference values determined following the approach of 
\citet{1967ApJ...148..217W}, for $\bar{B}_{r,0}= 30\un{G}$.
The solar rotation rate is taken to be $\Omega_\odot= 
2.9\cdot10^{-6}\un{s^{-1}}$.}
\label{wdrefvalues}
\centering
\begin{tabular}{rlccc}
\hline \hline
& & $\Omega_\odot$ & $8\un{\Omega_\odot}$ & $48\un{\Omega_\odot}$ \\
\hline
$(d\dot{M}/d\theta)\dw{WD}$ & [10$^{11}$ g s$^{-1}$ rad$^{-1}$] & 
3 & 7 & 91 \\
$F\dw{M,WD}$ & [10$^{10}$ g s$^{-1}$ sr$^{-1}$] & 
5 & 10 & 145 \\
\hline
$(d\dot{J}/d\theta)\dw{WD}$ & [10$^{32}$ dyn cm$^2$ rad$^{-1}$] & 
9 & 23 & 101 \\
$L\dw{WD}$ & [10$^{20}$ cm$^2$ s$^{-1}$] & 
28 & 34 & 11 \\
\hline
$v\dw{\infty,WD}$ & [km s$^{-1}$] & 
816 & 2302 & 3298 \\
$p\dw{w,WD}$ & [10$^{18}$ dyn sr$^{-1}$] & 
4 & 24 & 478 \\
\hline
\end{tabular}
\end{table}

\subsection{Mass loss rate}
\begin{figure}
\includegraphics[width=\hsize]{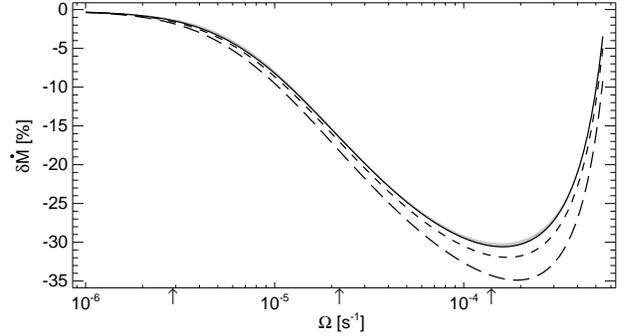}
\caption{ Relative deviations, $\delta \dot{M}$, of the
latitude-integrated mass loss rates from the WD reference values.
Constant Field (\emph{solid}); Polar Spot (\emph{long dashed});
Latitudinal Belt (\emph{short dashed}) distribution with
$\bar{B}_{r,0}= 30\un{G}$.
The thin \emph{grey shaded region} indicates the range of values for 
Constant Field distributions with field strengths between
$3-300\un{G}$.
Arrows mark rotation rates of the latitude-resolved cases shown in 
Fig.~\ref{mdotPL.fig}. }
\label{mdot.fig}
\end{figure}

\begin{figure}
\includegraphics[width=\hsize]{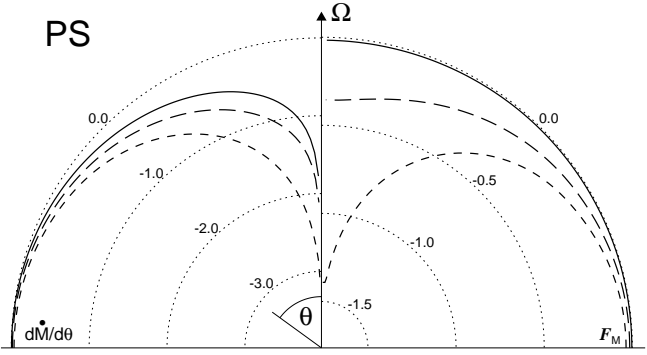}
\caption{ \textbf{Left quadrant}: Relative mass loss rate per latitude,
$\lg \left( (d\dot{M}/d\theta) / (d\dot{M}/d\theta)\dw{WD} \right)$.
\textbf{Right quadrant}: Relative mass flux, $\lg \left( F\dw{M} /
F\dw{M,WD} \right)$.
Polar Spot distribution with $\Omega= 2.9\cdot10^{-6}\un{s^{-1}}$ 
(\emph{solid}); $2.2\cdot10^{-5}\un{s^{-1}}$ (\emph{long dashed});
$1.4\cdot10^{-4}\un{s^{-1}}$ (\emph{short dashed}); the respective 
latitudinal profiles of the Latitudinal Belt and Constant Field 
distributions are qualitatively very similar and therefore not shown.
Note that with $r_0^2 \rho_0= \mathrm{const.}$ the relation
$F\dw{M}\propto v_{r,0}$ holds. }
\label{mdotPL.fig}
\end{figure}

With increasing rotation rate the latitude-integrated mass loss rates
fall short of the WD value, in the domain of rapid rotators up to about
$35\%$ (Fig.~\ref{mdot.fig}).
The major mass loss originates from the equatorial region, where the 
magneto-centrifugal driving of the wind is most efficient.
However, the mass loss rate per latitude, $d\dot{M} / d\theta$, 
decreases significantly with increasing latitude (relative to the WD 
values), depending on the rotation rate about two to three orders of 
magnitude (Fig.~\ref{mdotPL.fig}).
This strong decline is caused by the decreasing mass flux 
in combination with the latitudinal weighting function $\propto \sin 
\theta$, which takes the smaller surface fraction at higher latitudes
into account.
In the regime of slow stellar rotation the wind acceleration is caused
by the latitude-independent thermal driving, so that the mass flux is
virtually constant at all latitudes.
In this case the latitudinal variation of $d\dot{M} / d\theta$ is
dominated by the weighting function.
For larger rotation rates the latitudinal decrease of the mass flux 
strongly enhances the drop of the total mass loss rate.
Since the WD approach implies that the equatorial wind structure is 
representative for all latitudes, the generalisation of the equatorial
mass flux to higher latitudes results in an overestimation of the WD 
mass loss rate in comparison with the latitude-integrated values.

The deviations of the mass loss rates subject to latitude-dependent 
field distributions are larger than in the case of constant surface
fields, but retain the same functional behaviour, with largest 
deviations occurring at high rotation rates.
The Polar Spot yields mass loss rates up to $\sim 10\%$ smaller than 
the Constant Field distributions, whereas the Latitudinal Belt falls 
short only by a few percent.
High-latitude magnetic flux concentrations like a Polar Spot imply 
smaller field strengths at equatorial regions and consequently a drop 
of the mass flux and mass loss rate per latitude.
The Latitudinal Belt possesses at intermediate latitudes in contrast
sufficiently large field strengths and mass fluxes to sustain mass loss
rates similar to the case of a Constant Field.
Since non-equatorial latitudes contribute less to the overall mass loss
rate, this localised mass flux surplus can however not balance the
reduction of mass loss at the equator.

\subsection{Angular momentum loss rate}
\begin{figure}
\includegraphics[width=\hsize]{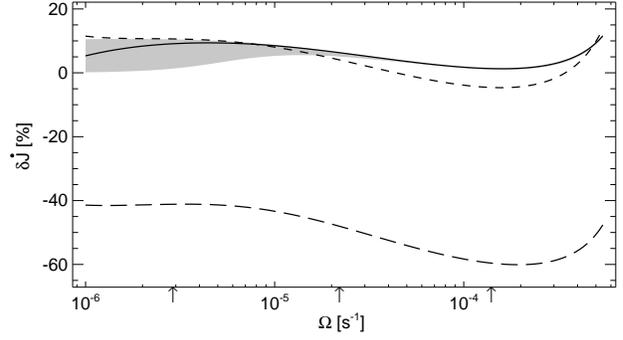}
\caption{ Relative deviations, $\delta \dot{J}$, of the 
latitude-integrated angular momentum loss rates from the WD reference
values.
Constant Field (\emph{solid}); Polar Spot (\emph{long dashed}); 
Latitudinal Belt (\emph{short dashed}) distribution with 
$\bar{B}_{r,0}= 30\un{G}$.
The \emph{grey shaded region} indicates the range of values for 
Constant Field distributions with field strengths between
$3-300\un{G}$.
Arrows mark rotation rates of the latitude-resolved cases shown in 
Fig.~\ref{jdotPL.fig}. }
\label{jdot.fig}
\end{figure}

The AM loss rates depend significantly on the stellar rotation rate and
magnetic field strength, but even stronger on the latitudinal field 
distribution (Fig.~\ref{jdot.fig}).
In the case of constant surface fields, the deviations of the
latitude-integrated AM loss rates from the WD values are at small
rotation rates moderate, up to about $10\%$, and field
strength-dependent.
In contrast, at large rotation rates the values become only marginal 
and independent of the field strength.
In the case of non-uniform field distributions, the integrated AM loss
rates can be significantly different \emph{despite of equal total 
amounts of open magnetic flux}.
In the case of a rapidly rotating star with a Polar Spot the total AM 
loss rate is about $60\%$ smaller than the WD value.
In the case of a Latitudinal Belt distribution it is about $10\%$
larger.
In contrast to the mass loss rate, where the deviations increase from 
marginal values at small rotation rates up to a maximum in the regime 
of fast rotation, the deviations of the AM loss rate are significant 
over the entire range of rotation rates.
The functional dependence on the stellar rotation is similar for all
field distributions.

\begin{figure}
\includegraphics[width=\hsize]{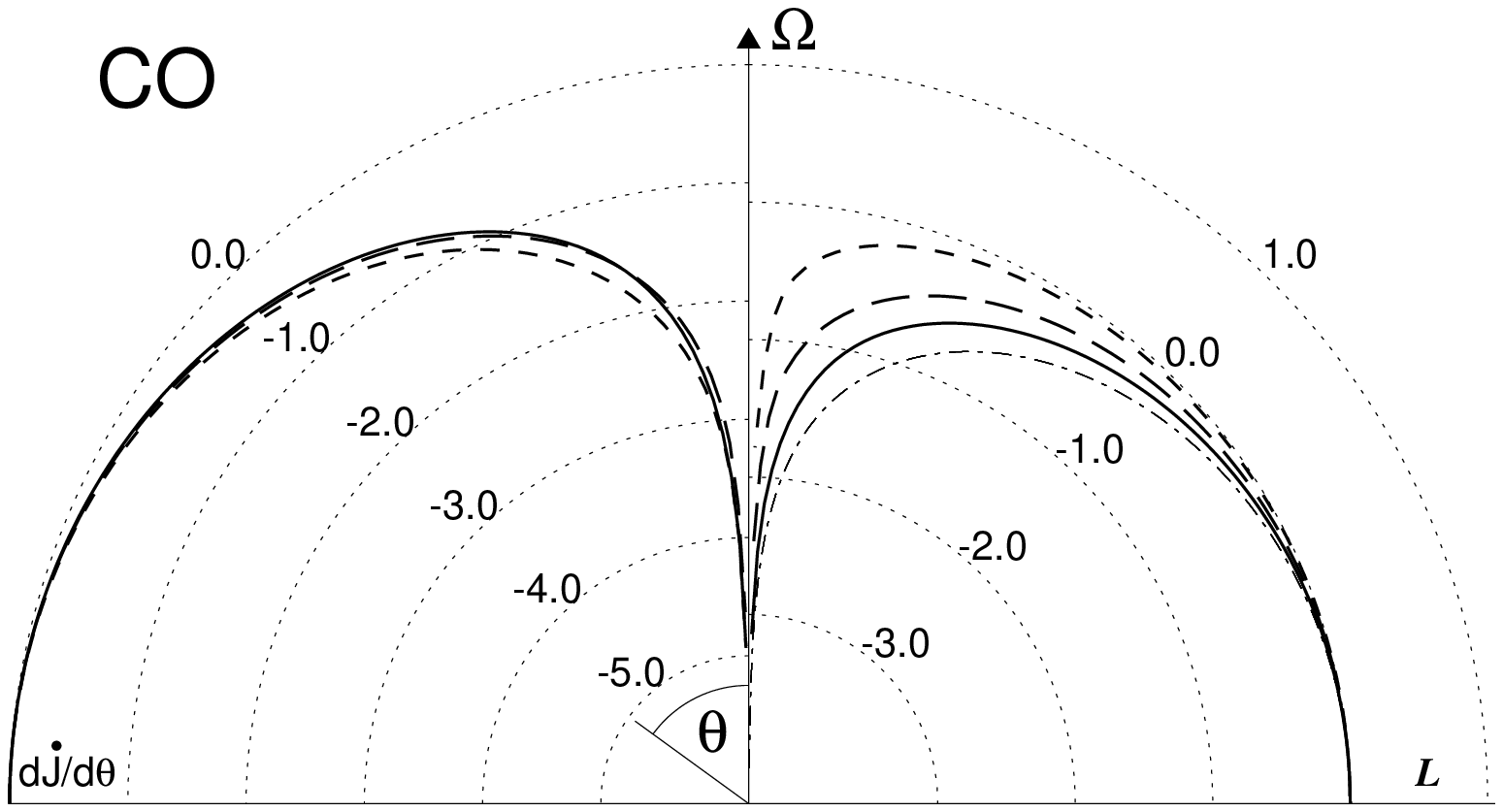}
\includegraphics[width=\hsize]{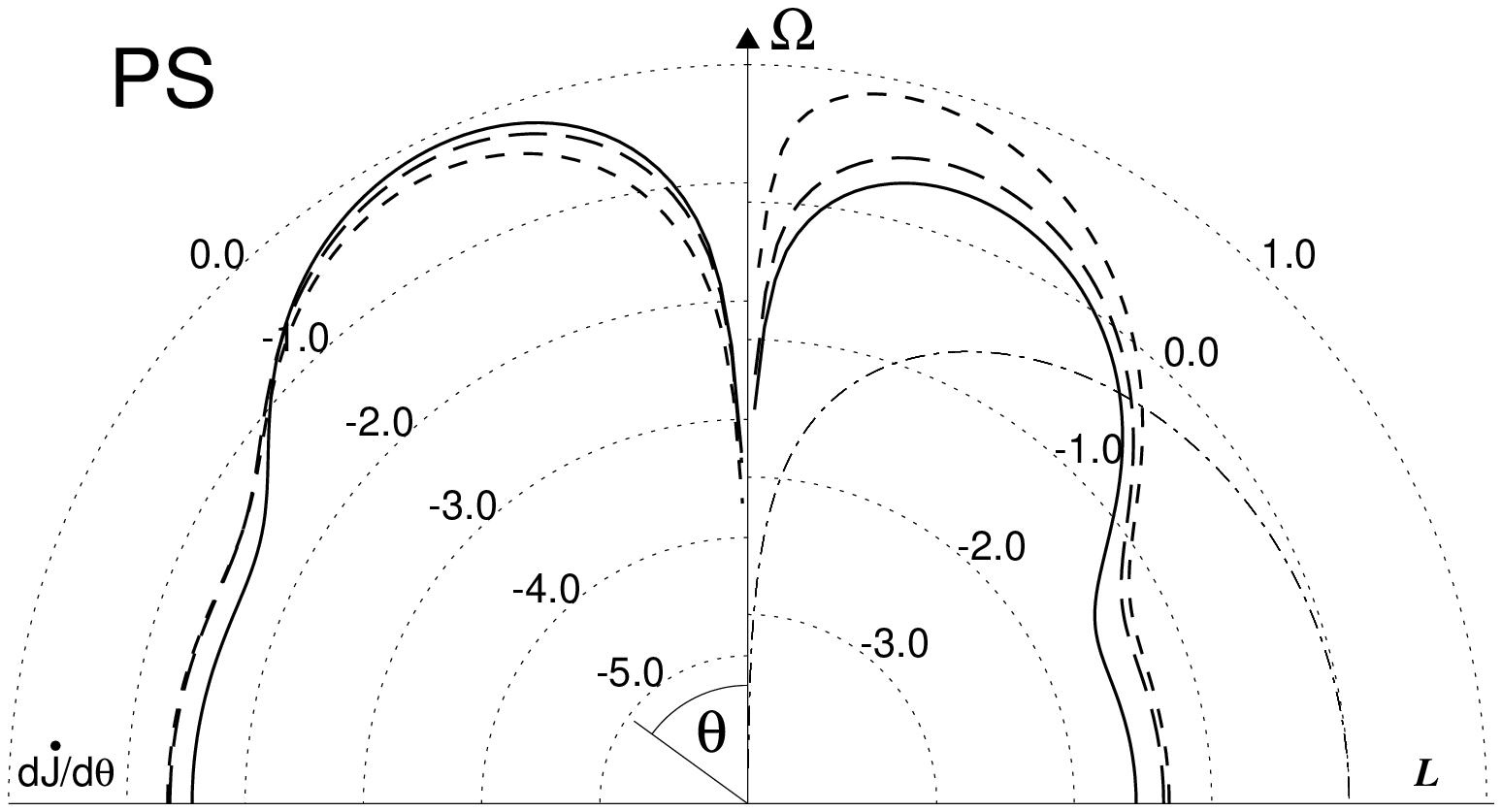}
\includegraphics[width=\hsize]{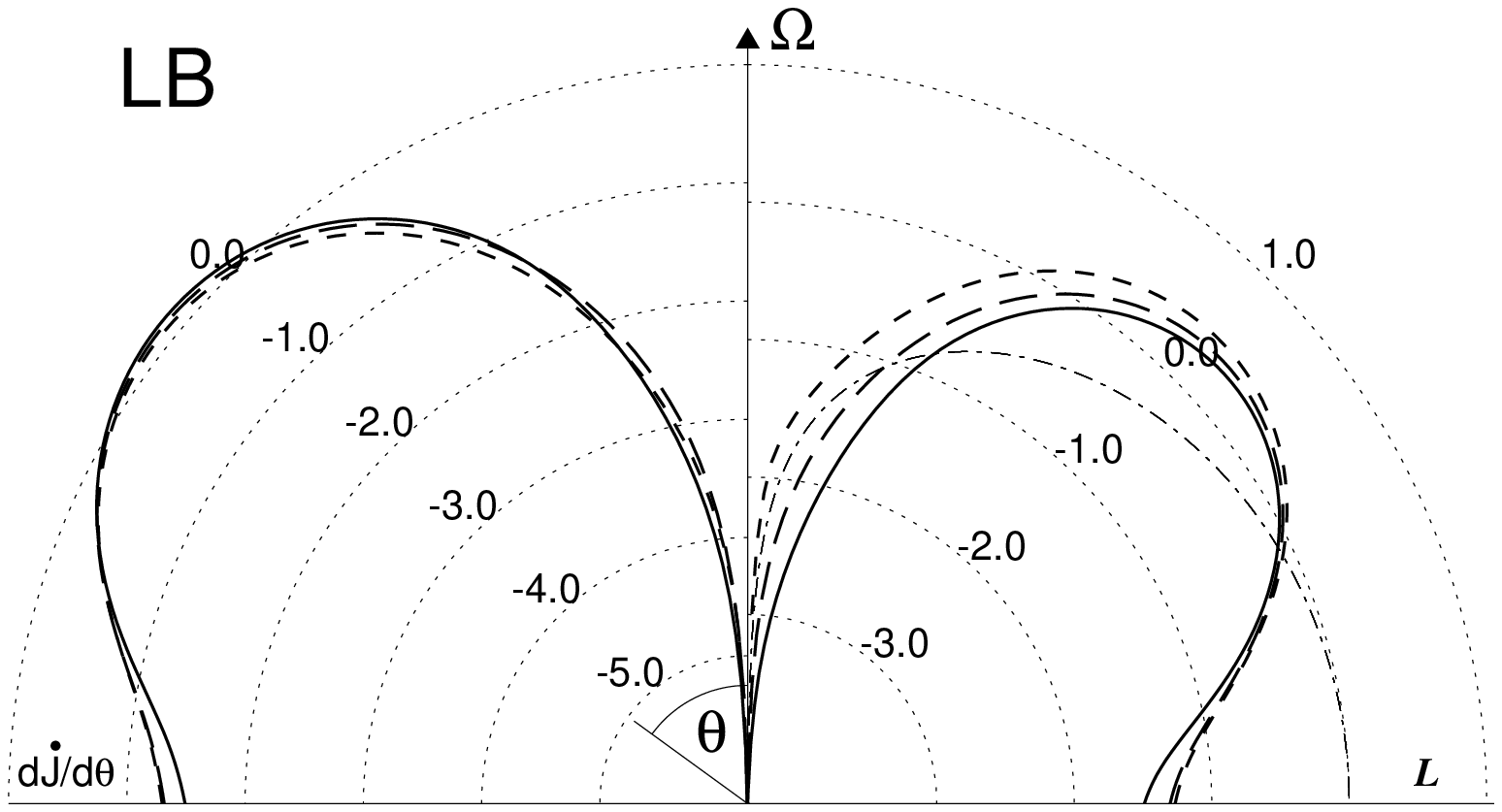}
\caption{ \textbf{Left quadrants}: Relative angular momentum loss rate
per latitude, $\lg \left[ (d\dot{J}/d\theta) /
(d\dot{J}/d\theta)\dw{WD} \right]$.
\textbf{Right quadrants}: Relative specific angular momentum, 
$\lg \left[ L / L\dw{WD} \right]$.
Constant Field (\emph{CO}); Polar Spot (\emph{PS}); Latitudinal Belt
(\emph{LB}) distribution with $\bar{B}_{r,0}= 30\un{G}$ and $\Omega=
2.9\cdot10^{-6} \un{s^{-1}}$ (\emph{solid});
$2.2\cdot10^{-5}\un{s^{-1}}$ (\emph{long dashed});
$1.4\cdot10^{-4}\un{s^{-1}}$ (\emph{short dashed}).
The \emph{dashed-dotted} lines show the weighting function $\propto 
\sin^2 \theta$; since $L\propto r\dw{A}^2 \sin^2 \theta$, the
differences between this line and the solid/dashed lines are caused by 
the latitudinal variation of the Alfv\'enic radius. }
\label{jdotPL.fig}
\end{figure}

The functional behaviour of the AM loss rate is caused by its combined
dependence on the mass loss rate (per latitude) and the specific AM,
the latter being very susceptible to field strength variations
(Fig.~\ref{jdotPL.fig}).
The specific AM is a latitudinally weighted function $\propto 
\sin^2 \theta$, due to its dependence on the \emph{axial} distance of 
the Alfv\'enic point.
The Alfv\'enic radius itself increases with latitude and decreases with
the rotation rate (cf.\ Fig.~\ref{wdref.fig}), which reflects the 
changing efficiency of the magneto-centrifugal driving.
In the case of uniform magnetic fields, the foreshortening of the 
geometric lever arm toward high latitudes is therefore countered by the 
concomitant increase of the Alfv\'enic radius, and a dependence of the 
specific AM on the rotation rate opposite to the one of the mass 
loss rate.
The combination of the latter two quantities results in a lateral 
compensation and therefore in similar $d\dot{J} / d\theta$-profiles 
(Fig.~\ref{jdotPL.fig}, CO).
Consequently, the WD approach underestimates the specific AM of 
non-equatorial regions by assigning the equatorial value of the 
Alfv\'enic radius to all latitudes. 
This underestimation is partly balanced by the overestimation of the 
mass loss rate, which eventually entails moderate deviations of the
latitude-integrated AM loss rates from the WD values.

A Polar Spot distribution yields the largest differences in the
latitudinal profile of the specific AM.
Although the magnetic field near the pole is very strong, its 
contribution to the AM loss rate (per latitude) is only average, since 
at high latitudes the mass loss rates are very small.
The reduction of the specific AM at low latitudes causes in 
contrast a much larger deficit, so that the total AM loss rate is far 
below the WD value.
The Latitudinal Belt distribution also shows a drop of the specific AM 
around the equator, but since its low-latitude field strengths are
larger than in the Polar Spot distribution the differences are in total
rather small.
The concentration of magnetic flux in a Latitudinal Belt implies high
magnetic field strengths at intermediate latitudes.
The large specific AM yields in combination with the substantial mass 
loss rates AM loss rates (per latitude), which are even
larger than in the reference case.

\subsection{Effective Alfv\'enic radius}
\begin{figure}
\includegraphics[width=\hsize]{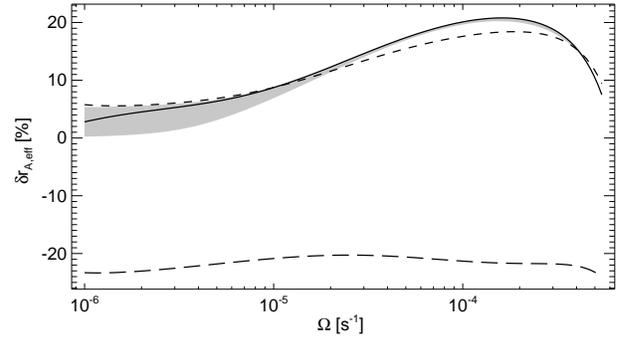}
\caption{ Relative deviations, $\delta r\dw{a,eff}$, of the effective
Alfv\'enic radii from the WD reference values.
Constant Field (\emph{solid}); Polar Spot (\emph{long dashed}); 
Latitudinal Belt (\emph{long dashed}) distribution with $\bar{B}= 
30\un{G}$.
The \emph{grey shaded region} indicates the range of values for 
Constant Field distributions with field strengths between
$3-300\un{G}$. }
\label{effalfr.fig}
\end{figure}

For the cases considered here, the deviations of the effective 
Alfv\'enic radius from the WD values are within about $\pm 25\%$ 
(Fig.~\ref{effalfr.fig}).
For a Constant Field or Latitudinal Belt distribution the effective
Alfv\'enic radius is larger than the WD value; at small rotation rates 
the difference is only marginal, but in the case of rapidly rotating 
stars it reaches a maximum of $\sim 20\%$.
The Polar Spot distribution, in contrast, yields effective Alfv\'enic 
radii which are for all rotation rates nearly $25\%$ smaller than the
reference values.

\subsection{Terminal wind velocity and ram pressure}
\begin{figure}
\includegraphics[width=\hsize]{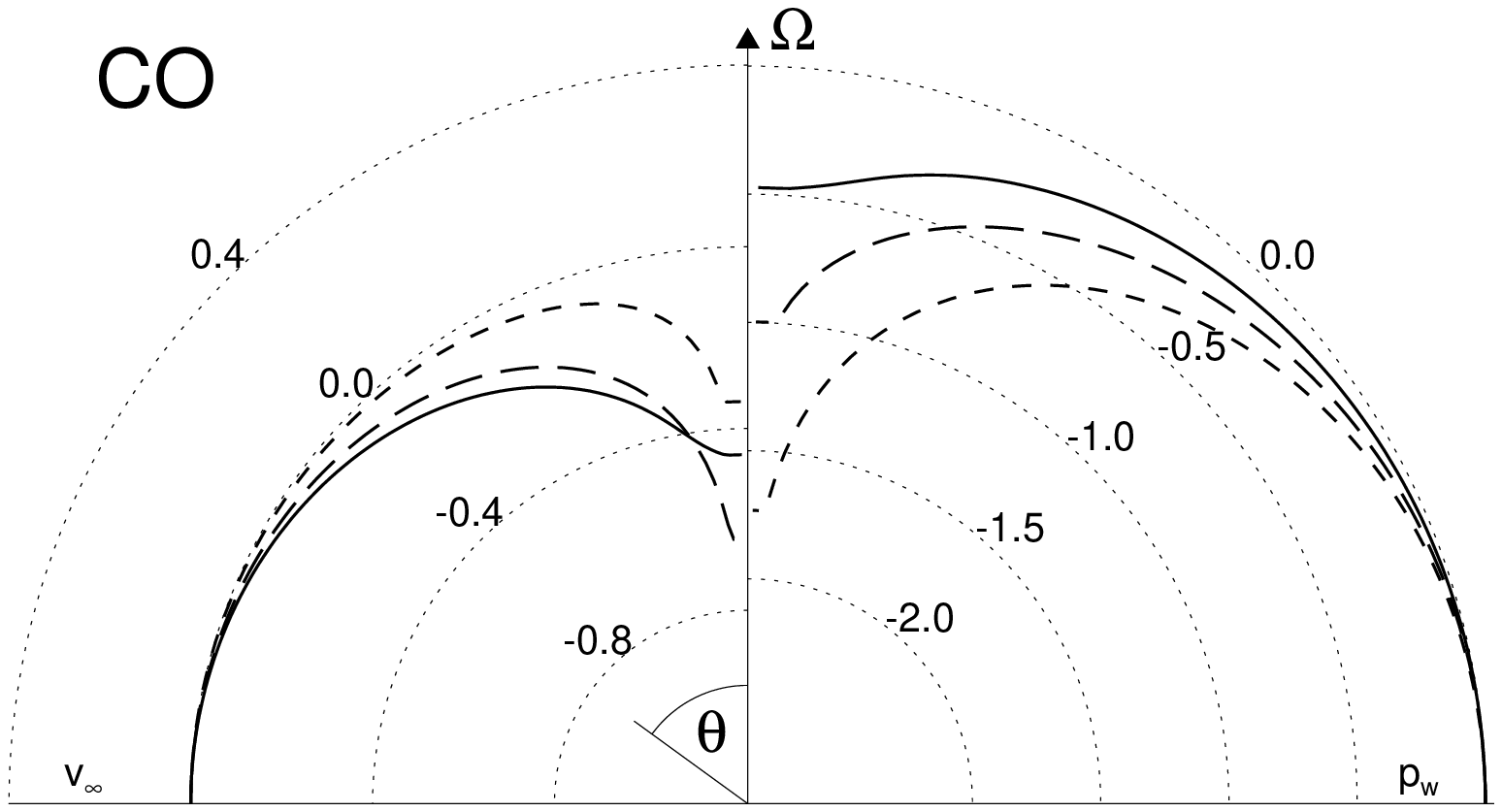}
\includegraphics[width=\hsize]{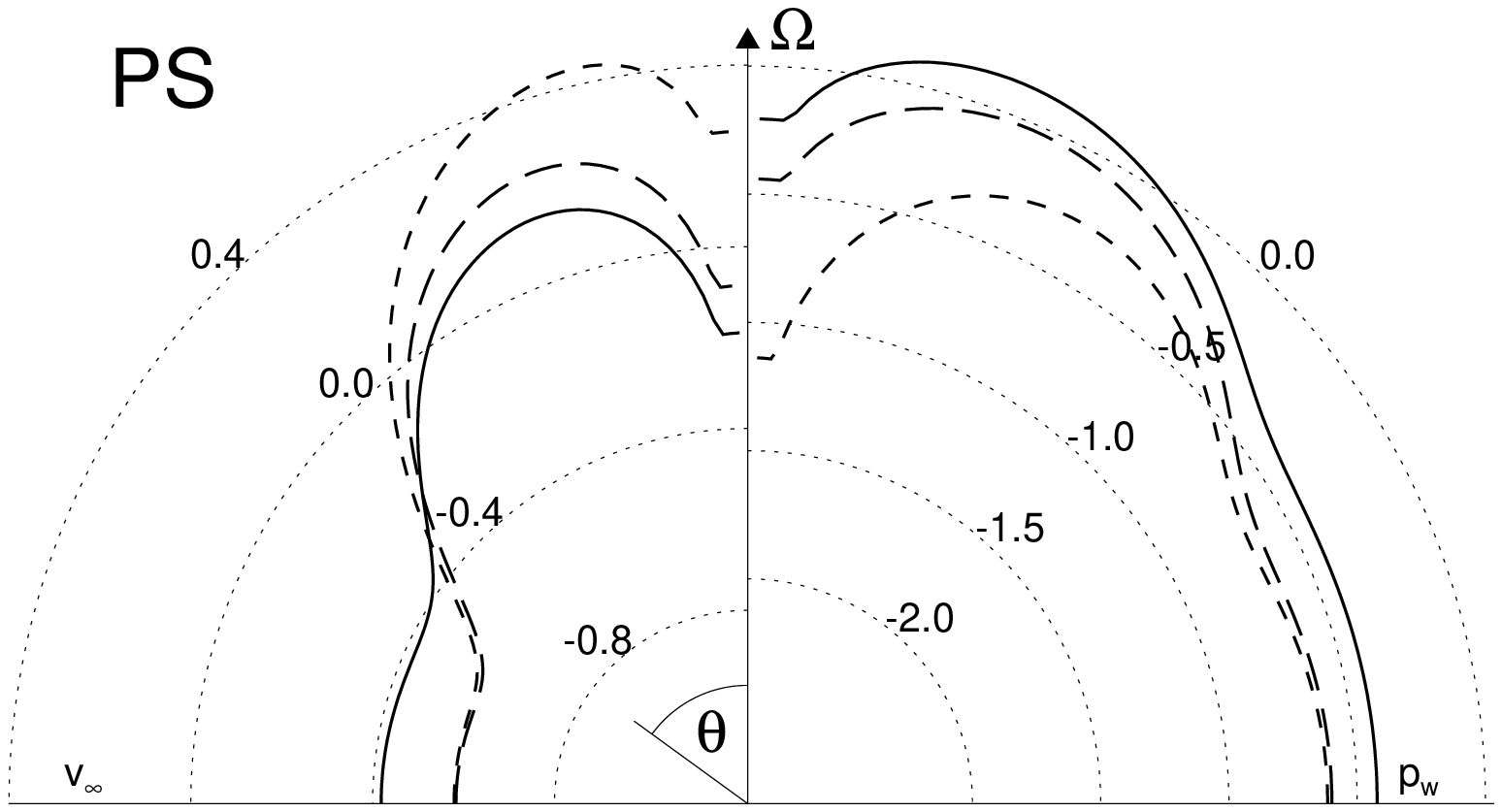}
\includegraphics[width=\hsize]{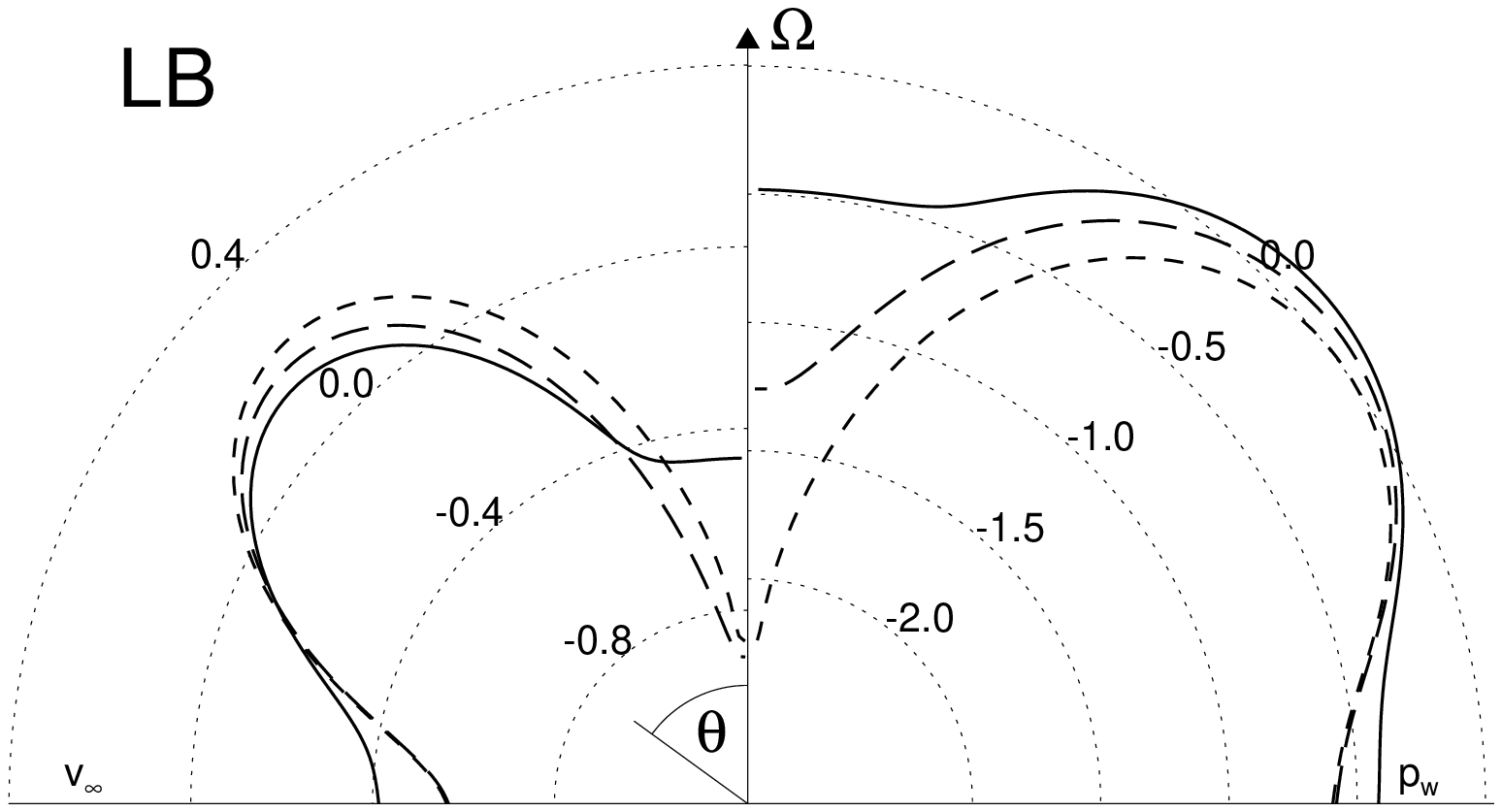}
\caption{ \textbf{Left quadrants}: Relative terminal wind velocity,
$\lg \left( v_\infty / v\dw{\infty,WD} \right)$.
\textbf{Right quadrants}: Relative wind ram pressure, $\lg \left( 
p\dw{w} / p\dw{w,WD} \right)$.
Constant Field (\emph{CO}); Polar Spot (\emph{PS}); Latitudinal Belt 
(\emph{LB}) distribution with $\bar{B}_{r,0}= 30\un{G}$ and 
$\Omega= 2.9\cdot10^{-6} \un{s^{-1}}$ (\emph{solid}); 
$2.2\cdot10^{-5}\un{s^{-1}}$ (\emph{long dashed});
$1.4\cdot10^{-4}\un{s^{-1}}$ (\emph{short dashed}). }
\label{ramp.fig}
\end{figure}

The ram pressure, $p\dw{w}$, exerted by a magnetised stellar wind far
away from the star is approximately proportional to the product of its
mass flux (cf.\ Fig.~\ref{mdotPL.fig}) and terminal wind velocity,
$v_\infty$.
The latitudinal profile of the latter reflects the strong dependence of
the magneto-centrifugal driving on the magnetic field distribution and 
on the stellar rotation rate (Fig.~\ref{ramp.fig}).
For small field strengths and/or rotation rates this driving mechanism
is rather inefficient and the terminal wind velocity thus small.
In the case of slowly rotating stars, the latitudinal variation of the
ram pressure is found to be within a factor of $\sim3$ comparable with 
the WD value, $p\dw{w,WD}\propto \dot{M}\dw{WD} v\dw{\infty,WD}$.
At high rotation rates, however, the latitudinal variation depends 
significantly on the latitude-dependent magneto-centrifugal driving of
the wind, and consequently on the actual magnetic field distribution.
In case of a Polar Spot the wind acceleration at high latitudes is 
supported by strong magnetic fields.
There the large terminal velocities make up for smaller mass fluxes, so
that the resulting latitudinal variation of the ram pressure is about
one order of magnitude.
In case of the Latitudinal Belt distribution, for which the magnetic 
flux is concentrated at intermediate latitudes, the terminal wind
velocities at high latitudes become very small and the difference of
the ram pressure between intermediate and polar latitudes very large,
here over two orders of magnitude.

\section{Dependence on thermal wind properties}
\label{dotp}
The winds considered here are accelerated through thermal and 
magneto-centrifugal driving.
The latter mechanism is in contrast to the isotropic thermal driving 
inherently rotation- and latitude-dependent.
The fractional contribution of each driving mechanism to the overall
wind acceleration determines the emphasise of its latitude-dependent
character, which consequently changes with the stellar rotation rate.
A priori, the thermal driving complies better with the basic WD
assumption of a spherical symmetric wind geometry than the
latitude-dependent magneto-centrifugal driving.
To quantify the impact of the thermal wind properties on the previous 
results, each one of the thermal parameters is successively changed, 
while the others are kept constant;  the parameter ranges are $T_0= 
(1.7-2.5)\cdot10^6\un{K}$; $n_0= 10^{-7}-10^{-9}\un{cm^{-3}}$; $\Gamma=
1.05-1.175$.

\begin{figure*}
\includegraphics[width=\hsize]{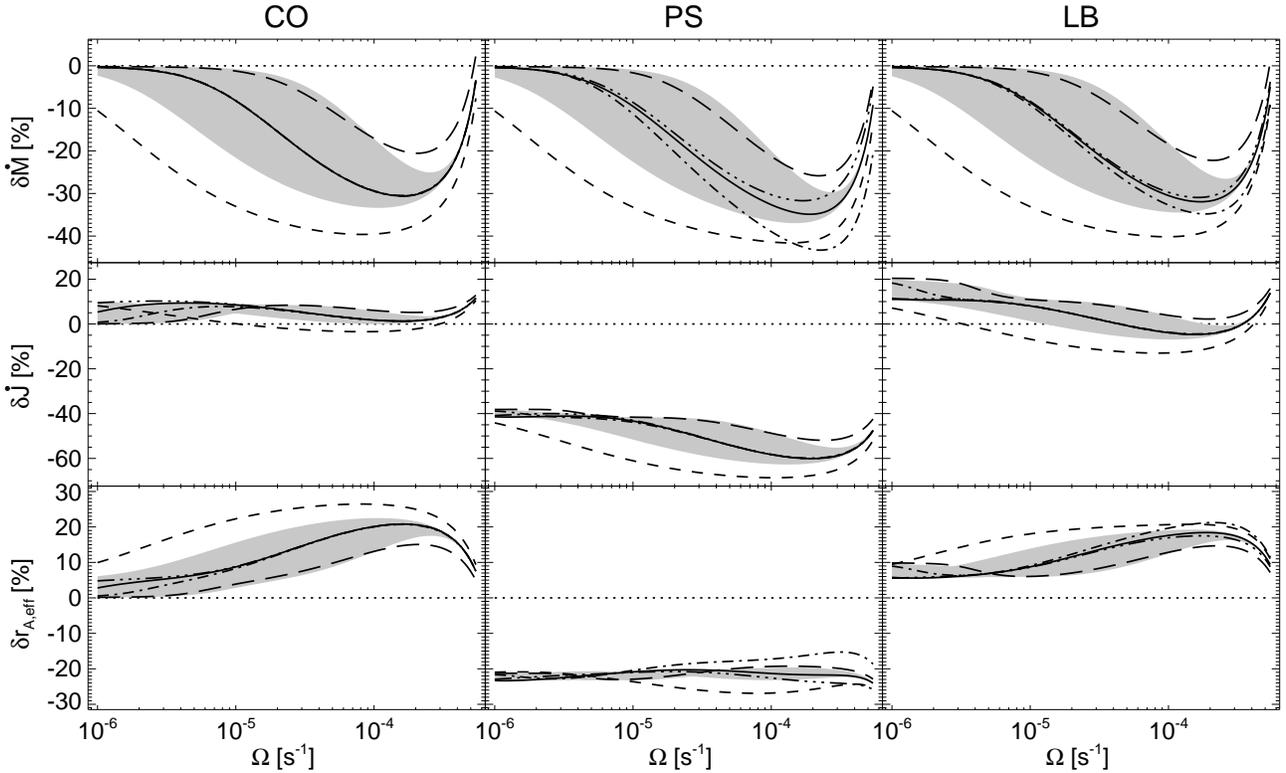}
\caption{ Relative deviations of the mass loss rate (\emph{top}), the
AM loss rate (\emph{middle}), and the effective Alfv\'enic radius
(\emph{bottom}) for the Constant Field (\emph{CO}); Polar Spot
(\emph{PS}); Latitudinal Belt (\emph{LB}) distribution.
The thermal properties at the base of the wind are $\Gamma= 1.15$ and
$T_0= 2\cdot10^6\un{K}\equiv T\dw{ref}, n_0= 10^8\un{cm^{-3}}\equiv 
n\dw{ref}$ (\emph{solid}); $T_0= 2.5\cdot10^6\un{K}, n_0= n\dw{ref}$ 
(\emph{long dashed}); $T_0= 1.7\cdot10^6\un{K}, n_0= n\dw{ref}$ 
(\emph{short dashed}); $T_0= T\dw{ref}, n_0= 10^9\un{cm^{-3}}$ 
(\emph{dashed-dotted}); $T_0= T\dw{ref}, n_0= 10^7\un{cm^{-3}}$ 
(\emph{dashed-triple dotted}).
The grey shaded regions indicate the range of values for polytropic 
indices $\Gamma= 1.05-1.175$ and $T\dw{ref}, n\dw{ref}$; the impact of
a lower polytropic index is thereby comparable with a higher wind 
temperature.
The (surface averaged) magnetic field strength is in all cases 
$\bar{B}_{r,0}= 30\un{G}$. }
\label{paradep.fig}
\end{figure*}

The mass loss rates of rapidly rotating stars, whose outflows are 
dominated by the magneto-centrifugal driving, are generally less
susceptible to changes of the thermal wind properties than those of
slow rotators.
Higher temperatures and/or stronger heating (i.e. smaller polytropic
indices) of the wind shift the emphasise of its acceleration from the
magneto-centrifugal to the thermal driving (Fig.~\ref{paradep.fig}, top
row).
The transition between the regimes of thermally controlled and 
rotationally controlled winds thus occurs at somewhat higher rotation 
rates.
Since the thermal driving is taken to be isotropic, the integrated mass
loss rate becomes in turn less susceptible to latitudinal variations of
the magnetic field, so that the relative deviations, $\delta \dot{M}$,
are smaller.
For cooler wind temperatures the effect is vice versa;  the mass loss 
rates are up to $40\%$ smaller than the WD reference values, with 
considerable deviations even at small rotation rates.
Tenuous winds with lower densities result in smaller deviations of the 
mass loss rate than more substantial outflows.
Thereby, magnetic field concentrations at high latitudes entail a 
larger susceptibility of the mass loss rate to variations of the wind 
density than field concentrations at lower latitudes.

The impact of the thermal wind properties on the AM loss rate is rather
complex and cannot easily be described in simple terms 
(Fig.~\ref{paradep.fig}, middle row).
The specific AM along an open field line depends on the axial 
distance of the location where the Alfv\'enic Mach number ($\propto 
v\sqrt{\rho}/B)$ is unity.
Therefore, the smaller the magnetic field strength, or the larger the 
wind density and/or flow velocity, the closer to the star the 
Alfv\'enic point will be located.
For higher wind temperatures the specific AM is therefore smaller.
But since this decrease is outbalanced by the concomitant increase of 
the mass loss rate, the resulting AM loss rates are usually larger than 
for cooler winds.
For the thermal wind parameters considered here, the $\delta
\dot{J}$-deviations have magnitudes roughly similar to the standard 
cases (thick solid lines in Fig.~\ref{paradep.fig}), although 
the individual functional behaviour can be quite different.
In particular at small rotation rates the deviations are susceptible to
the thermal wind properties, showing a rather non-linear behaviour.

Owing to its definition, Eq.~(\ref{defraeff}), the effective Alfv\'enic
radius reflects the functional dependence of both the mass and AM loss 
rates.
Since the deviation of the mass loss rate is typically larger than the
deviation of the AM loss rate, the functional dependence of $\delta 
r\dw{A,eff}$ is less complex than $\delta \dot{J}$; an exception is the
case of the Polar Spot distribution, for which the impact on the AM 
loss rate is very large.
The dependence on the wind density is generally rather small, whereas
cooler (hotter) winds are found to entail larger (smaller) deviations 
from the WD values.
The same is respectively true for the dependence on the wind heating.

\section{Discussion}
\label{disc}
The non-uniform field distributions considered here are motivated
through observations of rapidly rotating stars, which frequently
indicate the occurrence of magnetic fields up to polar latitudes.
Theoretical models explain these high-latitude features in terms of the
pre-eruptive poleward deflection of magnetic flux by the Coriolis force
\citep{1992A&A...264L..13S}, and/or its post-eruptive poleward
transport by meridional motions \citep{2001ApJ...551.1099S}.
The approach of \citet[WD]{1967ApJ...148..217W} in contrast implies a
split monopole (i.e.~spherically symmetric) geometry of the magnetic
field and wind structure, which results in an overestimation of the
mass flux and an underestimation of the specific AM at high latitudes.
The latitude-integrated mass and AM loss rates subject to non-uniform 
surface fields show consequently large deviations from the WD approach.

The deviations of the mass loss rate are largest (up to $30-35\%$) for
rapidly rotating stars, since in this regime the dominance of the 
latitude-dependent magneto-centrifugal driving causes large asymmetries
in the mass flux between the equator and the pole.
The dependence of this quantity on the field strength is small, so that
a redistribution of magnetic flux from equatorial to intermediate or
high latitudes hardly changes its latitudinal pattern.
The mutual deviations of the mass loss rates between different 
non-uniform flux distributions are here $\lesssim 10\%$. 

Since the functional dependencies of the mass flux and of the specific 
AM on the rotation rate follow opposite trends, their combined
effects partly balance, which makes the deviation of the AM loss rate
from the WD value nearly rotation-independent.
Non-uniform fields, on the other side, impose considerable variations 
on the latitudinal distribution of the specific AM, which are 
inherited by the AM loss rate causing deviations between $-60\%$ and
$10\%$.
Typically, the stronger the concentration of magnetic flux at higher 
latitudes, the smaller the AM loss rate.
However, for flux concentrations in a Latitudinal Belt at intermediate 
latitudes the AM loss rate is found to be \emph{larger} than the WD 
value, because the higher magnetic field strengths outbalance the 
foreshortening of the geometric lever arm.

The effective Alfv\'enic radius reflects the individual properties of
the mass and AM loss rates; its deviation form the WD value are here 
found to be within about $\pm 25\%$.
Owing to the quadratic dependence on the Alfv\'enic radius, the AM loss
rate can thus deviate up to $\pm 50\%$ from the WD values, if it is 
determined in terms of an empirical mass loss rate.
The largest uncertainties in the AM loss rate caused by disregarding 
the actual non-uniformity of surface magnetic fields are likely to 
occur in the case of rapid stellar rotation, when magnetic flux is 
gathering at high latitudes.

Both higher wind temperatures and densities are found to increase the 
mass loss rate, in agreement with previous results
\citep{1968MNRAS.138..359M, 1976ApJ...206..768Y}.
With increasing wind temperature (or heating) the isotropic thermal 
driving becomes stronger, and the importance of the magneto-centrifugal
driving and of non-uniform field distributions weaker.
The transition from thermally to rotationally controlled winds is thus
shifted to higher stellar rotation rates.
Vice versa, the mass loss rates of cool winds are more susceptible to
non-uniformities of the surface magnetic field.
The AM loss rates as well as the effective Alfv\'enic radii show a more
complex dependence on the thermal wind properties.
The order of magnitude of their deviations from the respective WD 
values remains roughly unaffected, but the functional behaviours change
in complex ways.
In accordance with the solar paradigm, high coronal temperatures are
ascribed to enhanced magnetic activity, which is found to increase with
the stellar rotation rate \citep{2001MNRAS.326..877M}.
Consequently, one has to assume that in rapidly rotating stars the 
structuring of stellar winds arising from the latitude-dependent 
magneto-centrifugal driving and from non-uniform surface magnetic
fields is diluted by high wind temperatures, which in turn fortify the
isotropic thermal driving.

In this work, the range of magnetic field strengths of the non-uniform 
field distributions is bracketed through the low- and high-field 
strength cases of uniform surface fields.
However, as mentioned above, their individual loss rates are
significantly different.
According to this, neither the surface averaged, nor the maximum, nor
the minimum field strength enable an accurate characterisation of
stellar magnetic fields.
In contrast, the AM loss rates of different non-uniform field
distributions can cover a large range of values, although their total
magnetic flux through the stellar surface is the same.
This has important consequences for studies considering the rotational
history of stars and the distribution of their rotation rates.
By investigating the impact of non-uniform surface magnetic fields on
the rotational evolution of stars, \citet{2005srestnsdoomf} show that
concentrations of magnetic flux at high latitudes cause rotational
histories which deviate up to $200\%$ from those obtained following the
WD approach.
We find that this impact is large enough to mimic deviations of the 
dynamo efficiency from linearity up to about $40\%$, and dynamo
saturation limits at about $35$ times the solar rotation rate
\citep[see also][]{1997A&A...325.1039S}.

\citet{1998ApJ...492..788W} and \citet{2002ApJ...574..412W} analysed 
the Ly$\alpha$ absorption from the hot \ion{H}{I} wall surrounding 
astrospheres to determine mass loss rates of solar-like stars.
The amount of absorption scales roughly with the square root of the
ram pressure of the associated (magnetic) stellar wind, which is in
turn a function of the mass loss and wind velocity.
To convert the ram pressure into mass loss rates they assume a unique 
wind velocity ($v\dw{w}= 400\un{km/s}$), taken to be isotropic and 
independent of the stellar rotation rate.
Based on their results they determine an empirical relation between the
mass lass rates of cool stars and their X-ray surface flux.
In the framework of the present wind model the latitudinal profiles of
the terminal wind velocity and ram pressure are however found to depend
considerably on the stellar rotation rate.
For slow rotators the ram pressure varies within a factor of $\sim 3$ 
with latitude.
For rapidly rotating stars, in contrast, the latitudinal variation can
rise to over two orders of magnitude.
The latitudinal profile of the ram pressure in addition depends
strongly on the non-uniformity of the surface magnetic field.
If most of the open magnetic flux is located near the poles, then the
strong magnetic fields there sustain the magneto-centrifugal driving at
high latitudes despite the less efficient centrifugal forces close to
the stellar axis of rotation; in this case the latitudinal variation of
the ram pressure is about one order of magnitude.
If most of the stellar magnetic flux is concentrated at intermediate
latitudes, then the polar magnetic field and wind acceleration will in
contrast be very weak.
Since the strong magneto-centrifugal driving at intermediate latitudes
entails high mass fluxes and wind velocities, the difference of the ram
pressure with respect to the polar value is over two orders of
magnitude.
The wind ram pressure derived along a single direction is consequently
not representative for the entire stellar surface.
If rotational and magnetic particularities of the star are not taken 
into account, the determined mass loss rates (and derived empirical
relationships) can be subject to larger uncertainties.
Since the difference of the latitudinal variation between a Polar Spot
and a Latitudinal Belt distribution is about one and half magnitudes, a
considerable reduction of these uncertainties could be achieved by
verifying which of the two cases describes the actual magnetic field 
distributions of rapidly rotating stars more accurately.
But for this, the yet small number of ZDI observations and field 
extrapolations has to be increased considerably.
The X-ray luminosities of the stars considered by 
\citet{2002ApJ...574..412W} are below the X-ray saturation limit 
\citep{1997ApJ...479..776S} and the objects thus likely moderate
rotators with rotation periods longer than about $2\un{d}$.
The possible inaccuracies of the ram pressure should accordingly be 
less than an order of magnitude, and the uncertainty of the mass loss 
rates due to the latitudinal variation less then a factor of $\sim 3$.
For stars of moderate rotation rate the linear empirical relationship 
between the stellar X-ray emission and mass loss rate should therefore
hardly be affected.
Due to the increasing latitudinal variation with rotation rate, the 
scatter of future observations of more rapidly rotating stars (with 
higher X-ray fluxes) around the predicted linear relationship is
however expected to increase.
The conclusion of \citeauthor{2002ApJ...574..412W}, that the mass loss
rates of younger and more rapidly rotating stars is up to several
orders of magnitude larger than in the case of the Sun, confirms the 
mass loss rates of rapid rotators determined in the framework of the WD
approach.

In their investigation of properties of stellar magnetic fluxes and
magnetised winds, \citet{2003ApJ...590..493S} combine observational 
stellar data in a chain of arguments, which involves the functional
dependence of the (effective) Alfv\'enic radius on the stellar rotation
rate.
Since in the framework of the present wind model, the difference
between the effective Alfv\'enic radius and its respective WD value
hardly depends on the rotation rate, the impact of non-uniform field 
distributions on their work is expected to be marginal.
In particular since their analyses is constrained to moderately active,
slowly rotating stars, whose magnetic flux is typically located at low
latitudes, in which case the effective Alfv\'enic radii are within 
$\sim 10\%$ comparable to the WD value.
However, as in more rapidly rotating stars magnetic flux is also found
at intermediate and high latitudes, the transition between different
latitude-dependent field distributions implies an additional dependence
of the Alfv\'enic radius on the rotation rate, which would have to be
taken into account.
The explicit functional dependence has to be determined observationally
\citep[e.g.][]{2002AN....323..309S} or through numerical simulations 
\citep[e.g.][]{1996A&A...314..503S, 2000A&A...355.1087G}.

In addition to the surface magnetic field distribution, the loss rates
of magnetised winds depend significantly on coronal magnetic field 
topologies. 
Theoretical models indicate that the mass and AM loss rates of a 
multi-component corona consisting of both wind and dead zones (i.e.,
open and closed field regions, respectively) are considerably smaller
than according to the WD model, since in addition to the confinement of
coronal plasma in magnetic loops dead zones also alter the flow
structure in adjacent wind regions \citep{1974SoPh...34..231P, 
1987MNRAS.226...57M}.
Observations in the case of the Sun however indicate that the mass loss 
rate arising from the slow solar wind, traced back to closed equatorial
field structures, is within a factor of two comparable with the mass 
loss rate of the fast solar wind, which originates from open coronal 
holes \citep{1998csss...10..131W}.

Simplified approaches to the wind model like the one used here
typically ignore the non-linear influence of the trans-field component
so that the resulting loss rates are subject to inaccuracies.
A comparison between in situ measurements of the solar wind by the 
\emph{Helios} spacecrafts and respective WD-based estimates however 
showed that this basic model is an adequate description of the physical 
interaction coupling the rotation of a star with its associated wind, 
insofar as simple magnetic stresses are taken as the principal physical
mechanism \citep{1983ApJ...271..335P}; for a comparison between 
out-of-eclipse measurements by the \emph{Ulysses} spacecraft with a 
self-consistent model approach see, for example,
\citet{2005A&A...432..687S}.
Furthermore, investigations of the rotational evolution of stars, which
make use of magnetic braking timescales estimated according to WD-like
wind models, yield rotational histories which are consistent with 
observed distributions of stellar rotation rates in young open clusters
of different age \citep[e.g.][]{1997A&A...325.1039S}.
Whereas multi-dimensional MHD-simulations solve the trans-field 
component of the equation of motion self-consistently 
\citep[e.g.,][]{1985A&A...152..121S, 1990CoPhyRep.12.247S,
2000ApJ...530.1036K}, in the present work the poloidal component of the
magnetic and velocity field are taken to be radial.
The numerical simulations show that at large distances from the star
magnetic field lines do not remain radial but tend to be deflected away
from the equatorial plane to higher latitudes, leading to a collimation
of the wind around the pole \citep{2000A&A...356..989T,
2000MNRAS.318..250O}.
The influence of this phenomenon on the mass and AM loss rates is not 
clear, since respective investigations are focused on the detailed wind 
structure of individual illustrative cases like the Sun.
The combination of multi-dimensional MHD-simulations with more 
appropriate field distributions is desirable, although the large 
numerical demand makes such modelling more challenging for stellar 
rotational evolution studies, where a large number of 
simulations is required to determine the relative impact of the 
evolving stellar structure, internal AM transport, dynamo efficiency, 
dynamo saturation, and surface magnetic field distributions on the
stellar rotation \citep[e.g.][]{2005srestnsdoomf}.

\section{Conclusion}
\label{conc}
The frequently observed non-uniform surface magnetic fields of active
stars have a remarkable influence on stellar mass and AM loss rates.
Concentrations of magnetic flux at high latitudes in the form of polar
spots reduce the AM loss rate and effective Alfv\'enic radius up to 
$60\%$ and $25\%$, respectively, with respect to the approach of 
\citet{1967ApJ...148..217W}.
The large deviations make them important ingredients for studies 
considering the rotational evolution of stars, implying considerable 
consequences for the distribution of stellar rotation rates.
The gathering of open magnetic flux in the form of a Latitudinal Belt 
has in contrast a strong impact on the error margins of empirical 
stellar mass loss rates derived using the ram pressure of stellar
winds.
In particular in the regime of rapidly rotating stars, the strong 
latitude-dependent magneto-centrifugal driving causes latitudinal 
variations of the terminal wind velocity and ram pressure which span
more than two orders of magnitude.
These examples show that the non-uniformity of surface magnetic fields
represents an essential stellar property, whose neglect entails large 
uncertainties of observational and theoretical results.

The loss rates and effective Alfv\'enic radii resulting from different
non-uniform surface magnetic fields are found to cover a large range of
values, although the total magnetic flux of each field distribution is
the same.
This shows that the classification of stellar magnetic fields through a
single, allegedly characteristic, field strength (e.g. the peak- or
surface-averaged field strength) is considered to be inapplicable.
Instead, possible non-uniformities of the surface magnetic fields have
to be taken into account, based either on empirical relationships or in
the framework of theoretical constraints.
In view of the importance of the actual location of open magnetic flux,
this makes an increase of the yet small number of combined Zeeman-DI
observations and reconstructions of global stellar magnetic field
topologies highly eligible.

\begin{acknowledgements}
The author would like to thank M Jardine and S Jeffers for valuable 
discussions and comments, and the anonymous referee for suggestions,
which helped to improve the paper.
This research was funded by a PPARC standard grand
(PPA/G/S/2001/00144).
\end{acknowledgements}

\bibliographystyle{aa}
\bibliography{}

\appendix

\section{Polytropic wind model}
\label{powi}
The stellar wind is taken to be an ideal plasma with vanishing 
viscosity and infinite conductivity.
Its structure is determined by the stationary equation of motion, and
subject to a polytropic relation between the thermal pressure, $p$,
and density, $\rho$.
The polytropic index, $\Gamma$, quantifies the heating of the wind; a
value close to one implies a nearly isothermal wind flow and a value 
close to the ratio of specific heats an almost adiabatic expansion.

\subsection{Basic relations and invariants}
In a reference system co-rotating with the stellar surface a stationary
plasma flow obeys the relation $\vec{v}\dw{co}= \kappa \vec{B} / \rho$,
where $\vec{v}\dw{co}$ is the flow velocity, $\vec{B}$ the magnetic
field, and $\kappa$ the ratio between the mass flux and the magnetic
flux; along a magnetic field line the latter is constant.
The flow velocity in the rest frame of reference is given by
\begin{equation}
\vec{v}
= 
\vec{v}\dw{co} + \vec{\Omega} \times \vec{r}
=
\kappa \frac{\vec{B}}{\rho} + \vec{\Omega} \times \vec{r}
\ ,
\label{defvis}
\end{equation}
where $\vec{r}$ is the vector pointing from the centre of the star 
to the location of a gas element, and $\vec{\Omega}$ a vector parallel 
to the stellar rotation axis, whose modulus, $\Omega$, is the stellar 
rotation rate.
The wind is assumed to be axi-symmetric with respect to the rotation
axis, $\vec{e}_\Omega= \vec{e}_z$, that is all derivatives in
azimuthal direction, $\vec{e}_\phi$, vanish.

Owing to the rotational symmetry, the azimuthal component of the 
equation of motion can be written in the form
\begin{equation}
\vec{v} \cdot \nabla \left( v_\phi \varpi \right)
=
\frac{1}{4 \pi \rho} \vec{B} \cdot \nabla \left( B_\phi \varpi \right)
\ ,
\label{azcoeom}
\end{equation}
with $\varpi = r \sin \theta$.
Substituting Eq.~(\ref{defvis}) into Eq.~(\ref{azcoeom}) yields that 
the \emph{specific} angular momentum (per unit mass),
\begin{equation}
L 
= 
v_\phi \varpi - \frac{1}{4 \pi \kappa} B_\phi \varpi 
\ ,
\label{deflco}
\end{equation}
of plasma escaping along an open magnetic field line is constant.
In addition to the common specific angular momentum of a moving gas
element, $\propto \vec{r}\times \vec{v}$, this quantity also comprises 
the tension of the magnetic field bent by the gas motion\footnote{
Considering the physical connection of Eq.~(\ref{deflco}) with the
vorticity and current density, \citet{2004POP.11.1.28G} use for the
quantity $L$ the nomenclature \emph{poloidal vorticity-current density
stream function} and the symbol $K$ to distinguish it from the specific
angular momentum of gas flows in hydrodynamics.}.
Following Eq.~(\ref{deflco}), the azimuthal flow velocity is written in
the form
\begin{equation}
v_\phi
=
\frac{ L - \frac{\rho}{4\pi\kappa^2} \varpi^2 \Omega}{ \varpi
\left( 1 - \frac{\rho}{4\pi\kappa^2} \right) }
\ .
\label{vphiwithL}
\end{equation}
For this expression to be finite at the Alfv\'enic point, 
$(\varpi\dw{A}, \rho\dw{A}= 4\pi \kappa^2)$, the specific angular
momentum has to be,
\begin{equation}
L= \Omega \varpi\dw{A}^2= \mathrm{const.}
\label{lcoconst}
\end{equation}
The Alfv\'enic point marks the location along a magnetic field line
where the Alfv\'enic Mach number is one (i.e. $v\dw{A} = B\dw{A} /
\sqrt{4\pi\rho\dw{A}}$), and the flow velocity changes from
sub-Alfv\'enic to super-Alfv\'enic values.

Since (in the rest frame of reference) the flow velocity of the wind is 
oblique to the magnetic field, the resulting Poynting flux leads to an
energy transfer,
\begin{equation}
\frac{1}{4\pi\rho} \vec{v} \cdot \left[ 
 \left( \nabla \times \vec{B} \right) \times \vec{B} 
\right]
=
\Omega\, \vec{v} \cdot \nabla \left( v_\phi \varpi \right)
\ ,
\label{magterm}
\end{equation}
from the stellar rotation into the gas motion via the magnetic field, 
which contributes to the wind acceleration in the form of a
magneto-centrifugal driving.
Taking Eq.~(\ref{magterm}) and the polytropic gas relation into
account, the energy function,
\begin{equation}
H
=
\frac{v^2}{2}
+
\frac{ \Gamma }{ \Gamma - 1 } \frac{p}{\rho}
+
\Psi
-
v_\phi \Omega \varpi 
\ ,
\label{egyfct}
\end{equation}
with $\Psi$ being the gravitational potential, has to be constant along
a magnetic field line, 
\begin{equation}
H= E= \mathrm{const.}
\label{egyconst}
\end{equation}
The four terms of Eq.~(\ref{egyfct}) are (in given order) the kinetic,
the thermal, the gravitational, and the magneto-centrifugal
contribution to the specific energy of the outflow.
The Eqs.~(\ref{egyfct}) and (\ref{egyconst}) follow from the component
of the equation of motion tangential to the wind flow and correspond
to the Bernoulli integral.
The two constants, Eq.~(\ref{lcoconst}) and (\ref{egyconst}), are a
direct consequence of the invariance of the problem against azimuthal
and temporal translations (i.e. axial symmetric and stationary flow).
For a more detailed description of the theory of magnetised winds see 
\citet{1999stma.book.....M, 1999isw..book.....L}.

A full treatment of the stellar wind problem requires the
self-consistent solution of its trans-field component, which determines
the poloidal field structure of the stellar wind.
This aspect involves the treatment of a non-linear partial differential
equation, which requires more sophisticated theoretical and/or 
numerical methods \citep[e.g.,][]{1985A&A...152..121S, 
1998MNRAS.298..777V, 2000ApJ...530.1036K}.
More detailed investigations show that at large distances from the
star, in the asymptotic regime, magnetic field lines are deflected 
toward the pole \citep{1989ApJ...347.1055H, 2000A&A...356..989T,
2000MNRAS.318..250O}.
In the present work the trans-field component is neglected and the
poloidal field component instead a priori taken to be radial
\citep[cf.][]{1976ApJ...206..768Y}.
Consequently, the field lines form spirals around the rotation axis of
the star, which are located on conical surfaces with constant opening
angle.
This approach is a simple generalisation of the
\citet{1967ApJ...148..217W} method in the sense that the individual
boundary conditions and properties of open field lines are explicitly
taken into account \emph{at all latitudes}, although their mutual
interaction is suppressed.
Due to the latter simplification the absolute values should however
only be considered as lowest-order estimates.

\subsection{Non-dimensional wind solutions}
The determination of wind solutions follows the method of 
\citet{1990CoPhyRep.12.247S}; see also 
\citet[Sect.~2]{1985A&A...152..121S}.
Using the non-dimensional variables\footnote{Since along a magnetic 
field line the mass flux per solid angle, $F= r^2 \rho v_r$, is
constant, the relation $x^2 y z= 1$ holds.
In contrast to \citet{1985A&A...152..121S}, here the flow velocity is
used as an independent variable instead of the density.}
\begin{eqnarray}
x= \frac{r}{r\dw{A}} \left(= \frac{\varpi}{\varpi\dw{A}} \right)
\quad , \qquad
y= \frac{\rho}{\rho\dw{A}}
\quad , \qquad
z= \frac{v_r}{v_{r,{\rm A}}}
\ ,
\label{xyzdef}
\end{eqnarray}
Eq.~(\ref{egyfct}) is written in dimensionless form,
\begin{eqnarray}
\lefteqn{
 \tilde{ H }
 =
 H \frac{r\dw{A}}{G M_*}
 =
 \frac{K_\beta}{2} z^2
 +
 \frac{K_\Omega}{2} \left[
  \left( \frac{x z \left( 1 - x^2 \right)}{1 - x^2 z} \right)^2 - x^2
 \right]
}
\nonumber \\
& & {}
-
\frac{1}{x}
+
\frac{K_T}{\Gamma - 1} \left( x^2 z \right)^{1-\Gamma}
\ ,
\label{egyfctxz}
\end{eqnarray}
with the non-dimensional parameters
\begin{equation}
K_T
=
\frac{\Gamma}{2 x\dw{Pa}} \frac{1}{y_0^{\Gamma - 1}}
\ , \quad
K_\Omega
=
\frac{1}{x_{\Delta g}^3} \ , \textrm{ and} \quad
K_\beta
=
\frac{x_0^4 y_0}{\beta_0 x\dw{Pa}}
\ .
\label{kdefs}
\end{equation}
Quantities with index '0' refer to values at the reference level,
$r_0$.
 
The parameter $K_\Omega$ is the only quantity that mediates a
dependence of the wind on both the stellar rotation rate and latitude.
In the equatorial plane the magneto-centrifugal driving of the stellar
wind is quantified through the co-rotation radius, $r\dw{co}=
\sqrt[3]{G M_\ast/ \Omega^2}$, which specifies the distance from the
star where the inward directed gravitation is balanced by the outward
directed centrifugal force.
Here, the magnetic field lines outside the equatorial plane are
situated on coni with constant opening angles.
Hence the latitude- and rotation-dependent radius
\begin{equation}
r_{\Delta g}
=
\sqrt[3]{ \frac{G M_\ast}{\Omega^2} } \sin^{-2/3} \theta
\label{rdeldef}
\end{equation}
denotes the point beyond which \emph{in radial direction},
$\vec{e}_r$, the centrifugal acceleration of the gas,
$\vec{a}\dw{c}$, outbalances the gravitational acceleration,
$\vec{g}$, that is $\left( \vec{g} + \vec{a}\dw{c} \right) \cdot 
\vec{e}_r = 0$.
The parameters $K_T$ and $K_\beta$ depend on the boundary conditions 
$x_0$ and $y_0= 1 / (x_0^2 z_0)$, determined at the reference radius, 
$r_0$.
There, the base temperature, $T_0$, of the wind defines the Parker
radius,
\begin{equation}
r\dw{Pa}
=
\frac{G M_*}{2 c\dw{I,0}^2}
=
\frac{G M_* \mu}{2 \Re} \frac{1}{T_0}
\ ,
\label{rpa}
\end{equation}
where $c\dw{I,0}$ is the (isothermal) sound speed, and $\mu$ the mean 
atomic weight of the plasma.
The dependence of the wind on the magnetic field strength enters 
through the parameter 
\begin{equation}
\beta_0
= 
\frac{8 \pi p_0}{B_0^2}
\ ,
\label{betadef}
\end{equation}
which denotes the ratio between the gas pressure and the magnetic 
pressure at the base of the wind.

Since $\tilde{\mathcal{H}}= \tilde{E}= \mathrm{const.}$, a
non-dimensional wind solution, $z(x)$, is represented in the
$(x,z)$-plane by an isoline.
Along this trajectory the total differential of the energy function,
\begin{equation}
d \tilde{H}
=
\left( \frac{\partial \tilde{H}}{\partial x} \right)_z d x
+
\left( \frac{\partial \tilde{H}}{\partial z} \right)_x d z
\ ,
\label{htitdif}
\end{equation}
vanishes, so that the variation of the flow velocity in radial
direction is determined by the differential equation
\begin{equation}
\frac{d z}{d x}
=
-
\frac{ \left( \frac{\partial \tilde{H}}{\partial x} \right)_z }{
\left( \frac{\partial \tilde{H}}{\partial z} \right)_x }
\ .
\label{dzdxdef}
\end{equation}
To prevent this derivative from becoming infinite at points where the
numerator vanishes, the denominator simultaneously has to be zero.
The condition
\begin{equation}
\left( \frac{\partial \tilde{H}}{\partial z} \right)_x
=
\left( \frac{\partial \tilde{H}}{\partial x} \right)_z
= 0
\label{criptsdef}
\end{equation}
defines two locations along the trajectory, which are usually referred 
to as the slow and fast critical points, $(x\dw{s},z\dw{s})$ and 
$(x\dw{f},z\dw{f})$, respectively.
Since both points are situated on the same energy level, the condition 
\begin{equation}
\tilde{\mathcal{H}} \left( x\dw{s},z\dw{s} \right)
= 
\tilde{\mathcal{H}} \left( x\dw{f},z\dw{f} \right) 
= 
\tilde{\mathcal{E}}
\label{eqegydef}
\end{equation}
holds.
The five non-linear algebraic Eqs.~(\ref{criptsdef}, for the slow and 
fast critical points) and (\ref{eqegydef}) are used to determine sets
of non-dimensional wind solutions $(x\dw{s}, z\dw{s}, x\dw{f}, z\dw{f},
K_\beta)$ as a function of the two remaining parameters $K_T$ and
$K_\Omega$.
These solutions are as yet independent of specific boundary conditions;
the only physical quantity explicitly entering Eq.~(\ref{egyfctxz}) 
is the polytropic index.
For a given value of $\Gamma$, a look-up table is pre-calculated to 
speed up the following determination of dimensional wind solutions, 
which depend on the boundary conditions at the reference level.

\subsection{Dimensional boundary conditions}
In addition to the stellar rotation rate, $\Omega$, a dimensional wind 
solution requires the definition of the temperature, $T_0$, the
density, $\rho_0$, and the radial magnetic field strength, $B_{r,0}$ at
the reference level, $r_0$.
The link between these dimensional boundary conditions and the 
dimensionless wind solutions above is established through a combination
of Eqs.~(\ref{kdefs}),
\begin{equation}
K_T K_\Omega^{\Gamma - 4/3} K_\beta^{\Gamma - 1}
=
\frac{\Gamma}{2} \left( \frac{r_{\Delta g}}{r_0} \right)^{4 - 3 \Gamma}
\left( \frac{r\dw{Pa}}{r_0} \right)^{-\Gamma} \beta_0^{1 - \Gamma}
\ ,
\label{paraco}
\end{equation}
to eliminate the yet unknown Alfv\'enic radius, $r\dw{A}$, and
density, $\rho\dw{A}$, as well as the energy relation
\begin{eqnarray}
\frac{ \tilde{ \mathcal{E} } }{K_\Omega^{1/3}}
=
-
\frac{r_{\Delta g}}{r_0} \left[
 1
 +
 \frac{1}{2} \left( \frac{r_{\Delta g}}{r_0} \right)^{-3}
 -
 \frac{1}{2} \frac{\Gamma}{\Gamma - 1} 
 \left( \frac{r\dw{Pa}}{r_0} \right)^{-1}
\right]
\ .
\label{egyco}
\end{eqnarray}
The latter is determined at the reference level, where the specific
kinetic energy of the wind flow is assumed to be negligible compared 
with the thermal and centrifugal energy. 
For a given set of boundary conditions, the right sides of
Eqs.~(\ref{paraco}) and (\ref{egyco}) are constant and the solution
found by an appropriate choice of $K_T$ and $K_\Omega$; note that
$K_\beta$ and $\tilde{\mathcal{E}}$ are both implicit functions of
these independent variables.
Once a non-dimensional wind solution has been found, the Alfv\'enic
radius and density are determined by back-substitution of the
parameters $K_T, K_\Omega$, and $K_\beta$.

\subsection{Terminal wind velocity}
The terminal wind velocity, $z_\infty$, is determined using 
Eq.~(\ref{egyfctxz}) in the limit of large distances from the star 
($x\rightarrow \infty$), which yields the cubic equation 
\begin{equation}
z^3 
- 
z\,
2 \frac{K_\Omega}{K_\beta} \left(1 + \frac{\tilde{E}}{K_\Omega}\right)
+
2 \frac{K_\Omega}{K_\beta} 
=
0
\ .
\label{tvelcubeq}
\end{equation}
The correct solution is subject to the constraint that even far away
from the star the wind velocity has still to increase, that is
\begin{equation}
\left. \frac{d z}{dx}\right|_{x\rightarrow \infty} 
\simeq 
\frac{2 z^2}{x} \frac{K_t y^{\Gamma-1}}{\left( z^3 K_\beta - K_\Omega
\right)}
> 0
\ .
\label{tvelinc}
\end{equation}
For a stationary outflow to exist, the terminal wind velocity has to 
exceed a latitude-dependent minimal flow velocity,
\begin{equation}
z_\infty > z\dw{M}= 
\sqrt[3]{\frac{K_\Omega}{K_\beta}}
= 
\frac{v\dw{M}}{v_{r,{\rm A}}} \sin^{2/3} \theta
\ ,
\label{defzmich}
\end{equation}
with $v\dw{M}= \sqrt[3]{\Omega^2 r\dw{A}^2 v_{r,{\rm A}}}$ being the
characteristic Michel velocity of the wind \citep{1969ApJ...158..727M}.
With a proper set of parameters $K_\Omega, K_\beta$, and $\tilde{E}$
obtained from Eqs.~(\ref{paraco}) and (\ref{egyco}), the terminal
velocity is determined from Eqs.~(\ref{tvelcubeq}) and
(\ref{defzmich}).

\label{lastpage}

\end{document}